\documentstyle[12pt]{article}
\makeatletter
\def\art{\@ifnextchar[{\eart}{\oart}}
\def\eart[#1]#2#3#4#5#6{{\rm #2}, {\em #3 \bf #4} {\rm (#6) #5} ({\em
#1})}
\def\hepart[#1]#2{{\rm #2, \em#1}}
\newcommand{\oart}[5]{{\rm #1}, {\em #2 \bf #3} {\rm (#5) #4}}
\newcounter{alphaequation}[equation]
\def\thealphaequation{\theequation\hbox to
0.6em{\hfil\alph{alphaequation}\hfil}}
\def\eqnsystem#1{
\def\@eqnnum{{\rm (\thealphaequation)}}
\def\@@eqncr{\let\@tempa\relax \ifcase\@eqcnt \def\@tempa{& & &} \or
  \def\@tempa{& &}\or \def\@tempa{&}\fi\@tempa
  \if@eqnsw\@eqnnum\refstepcounter{alphaequation}\fi
\global\@eqnswtrue\global\@eqcnt=0\cr}
\refstepcounter{equation} \let\@currentlabel\theequation \def\@tempb{#1}
\ifx\@tempb\empty\else\label{#1}\fi
\refstepcounter{alphaequation}
\let\@currentlabel\thealphaequation
\global\@eqnswtrue\global\@eqcnt=0 \tabskip\@centering\let\\=\@eqncr
$$\halign to \displaywidth\bgroup \@eqnsel\hskip\@centering
$\displaystyle\tabskip\z@{##}$&\global\@eqcnt\@ne
\hskip2\arraycolsep\hfil${##}$\hfil& \global\@eqcnt\tw@\hskip2\arraycolsep
$\displaystyle\tabskip\z@{##}$\hfil
\tabskip\@centering&\llap{##}\tabskip\z@\cr}

\def\endeqnsystem{\@@eqncr\egroup$$\global\@ignoretrue} \makeatother

\begin{document}

\def\lesssim{\mathrel{\mathpalette\vereq<}}
\def\gtrsim{\mathrel{\mathpalette\vereq>}}
\def\vereq#1#2{\lower3pt\vbox{\baselineskip1.5pt \lineskip1.5pt
\ialign{$\m@th#1\hfill##\hfil$\crcr#2\crcr\sim\crcr}}}
\makeatother

\newcommand{\NP}{Nucl. Phys.}
\newcommand{\PRL}{Phys. Rev. Lett.}
\newcommand{\PL}{Phys. Lett.}
\newcommand{\PR}{Phys. Rev.}

\newcommand{\rem}[1]{{\bf #1}}
\def\Red{}
\def\Black{}
\def\Blue{}

\renewcommand{\thefootnote}{\fnsymbol{footnote}}
\setcounter{footnote}{0}
\begin{titlepage}
\begin{center}

15 Aug.\ 98    \hfill    LBNL-42164\\
IFUP-TH/98-32  \hfill    SNS-PH/98-17\\
hep-ph/9808333 \hfill    UCB-PTH-98/49\\

\vskip .5in

{\Large \bf \Red
Textures for Atmospheric and Solar\\ Neutrino Oscillations
\footnote
{This work was supported in part by the U.S. 
Department of Energy under Contracts DE-AC03-76SF00098, in part by the 
National Science Foundation under grant PHY-95-14797
and  in part by the TMR Network under the EEC Contract
No. ERBFMRX - CT960090. }
}\Black

\vskip .50in

Riccardo Barbieri$^1$, Lawrence J. Hall$^2$ and A. Strumia$^3$

\vskip 0.2in
$^1$ 
{\em Scuola Normale Superiore, and INFN, sezione di Pisa\\
I-56126 Pisa, Italia}
\vskip 0.1in
$^2$
{\em Department of Physics and Lawrence Berkeley National Lab.,\\
     University of California, Berkeley, California 94720}
\vskip 0.1in
$^3$ 
{\em Dipartimento di fisica, Universit\`{a} di Pisa\\
and INFN, sezione di Pisa, I-56127 Pisa Italia}

\end{center}

\vskip .5in \Blue
\begin{abstract}
In theories with three light neutrinos, a complete list of five
zeroth-order textures for lepton mass matrices is found, which
naturally yield: $m_{e,\mu}=0, m_\tau \neq 0, \Delta m^2_\odot = 0$,
together with $\theta_{e \tau} = 0, \theta_{\mu \tau} \approx
1$. These textures provide suitable starting points for constructing
theories which account for both atmospheric and solar neutrino
fluxes. Using flavor symmetries, two schemes for such lepton masses
are found, and example constructions exhibited.
\end{abstract}\Black
\end{titlepage}

\renewcommand{\thepage}{\arabic{page}}
\setcounter{page}{1}
\renewcommand{\thefootnote}{\arabic{footnote}}
\setcounter{footnote}{0}

\paragraph{1}
The Super-Kamiokande Collaboration has provided strong evidence that 
atmospheric $\nu_\mu$ are depleted as they traverse the Earth~\cite{SK1,SK2}. 
For example, the up/down event ratio shows a $6 \sigma$ statistical
significance, and could only be explained by some systematic effect 
an order of magnitude larger than those already considered.
With two flavor $\nu_\mu \rightarrow \nu_\tau$ oscillations, the
mixing angle is large, $\sin^2 2 \theta > 0.82$, and $\Delta
m^2_{{\rm atm}}$ is within a factor of three of $1.5 \times 10^{-3}$ eV$^2$
at the 90\% C.L.~\cite{SK2}.

In this letter we study lepton masses in
theories with three light neutrinos.
With the assumptions given below, we find a 
complete list of five zeroth-order textures,
which can account both for this atmospheric data and for the
presence of solar neutrino oscillations at some frequency $\Delta
m^2_\odot \ll \Delta m^2_{{\rm atm}}$. Requiring the
textures to follow from symmetries, we are able to find only two 
zeroth-order schemes for the light lepton mass matrices which
give both the observed atmospheric and solar neutrino fluxes.

\paragraph{2}
Both charged and neutral lepton mass matrices involve small
dimensionless parameters: $m_e/m_\tau \ll m_\mu / m_\tau \ll 1$ and
$\Delta m^2_\odot / \Delta m^2_{{\rm atm}} \ll 1$. In the
flavor basis, the lepton mass matrices, $\bar{\ell}_L m_E \ell_R$ and 
$\nu_L^T m_{LL} \nu_L$, can be written as a perturbation
series in a set of small parameters $\epsilon$:
\begin{equation}
m_E = v \left( \lambda_E^{(0)} +  \lambda_E^{(1)}(\epsilon) + \cdots \right)
\label{eq:mE}
\end{equation}
\begin{equation}
m_{LL} = {v^2 \over M} \left( \lambda_\nu^{(0)} +
 \lambda_\nu^{(1)}(\epsilon) + \cdots \right)
\label{eq:mnu}
\end{equation}
where $v$ is the electroweak symmetry breaking scale and $M$ is some
new large mass scale, so that the $\lambda^{(i)}$ are dimensionless
contributions at $i$th order in perturbation theory.
The matrices $ \lambda_E^{(0)}$ and $ \lambda_\nu^{(0)}$ have entries
which are either zero or of order unity, and the possible forms
for these matrices, which we call zeroth-order textures, are the subject of 
this letter.\footnote{Our classification of textures does not assume any flavor
symmetry. However, perturbation series of the type~(\ref{eq:mE}) and 
(\ref{eq:mnu})
could result from a flavor symmetry which defines the flavor basis,
allows the terms $\lambda_{E,\nu}^{(0)}$, and leads to higher order
corrections as a power series in the small flavor symmetry breaking
parameters $\epsilon$.}

\smallskip

Diagonalization of $m_E$ and $m_{LL}$ gives both the lepton mass
eigenvalues and the leptonic mixing matrix $V$, defined by the charged
current interaction $\bar{\nu} V^T \gamma^\mu \ell \; W_\mu$.
Ignoring phases, which are irrelevant for our purposes, 
we define the three Euler angles by
$V = V_e^\dagger V_\nu = 
R_{23}(\theta_{23}) R_{13}(\theta_{13}) R_{12}(\theta_{12})$. 
The rotation matrices $V_{e,\nu}$, 
which diagonalize $m_{E,LL}$, are the same 
functions of the angles $\theta_{E,\nu ij}$ as $V$ is of $\theta_{ij}$.
We require that the diagonalization of  $\lambda_{E,\nu}^{(0)}$ leads
to zeroth-order eigenvalues
\begin{equation}
m_{E1}^{(0)} = m_{E2}^{(0)} = 0
\label{eq:memu}
\end{equation}
\begin{equation}
m_{E3}^{(0)} \simeq m_\tau
\label{eq:mtau}
\end{equation}
\begin{equation}
\Delta m_\odot^{2(0)} \equiv m_{\nu 1}^2 - m_{\nu 2}^2 = 0.
\label{eq:modot}
\end{equation}
In most cases $\lambda_{\nu}^{(0)}$ will lead to a
non-zero value for $\Delta m^2_{{\rm atm}}$; however, this is
not a strict requirement.

To diagonalize $\lambda_E^{(0)}$, we first rotate the right-handed 
charged leptons to obtain the form:
$$
\lambda_E^{(0)} = \pmatrix{0&0&C \cr 0&0&B \cr 0&0&A}.
$$
We diagonalize this form by choosing $\theta_{E12}^{(0)}=0$, which 
defines a basis for the light states, and determines 
$\theta_{E13,23}^{(0)}$ in terms of $A,B$ and $C$.
We further require that the diagonalization of $\lambda_{E,\nu}^{(0)}$
gives zeroth-order leptonic mixing angles
\begin{equation}
\theta_{23}^{(0)} \approx 1
\label{eq:theta23}
\end{equation}
and
\begin{equation}
\theta_{13}^{(0)} =0.
\label{eq:theta13}
\end{equation}
The requirement $\theta_{13}^{(0)} =0$ follows from 
recent fits to the Super-Kamiokande atmospheric data~\cite{barger},
which find $\theta_{13}$ less than approximately $20^\circ$, 
for a hierarchy of $\Delta m^2$
values such as we assume. Furthermore, for $\Delta m^2_{{\rm atm}} > 2 \cdot
10^{-3}$ eV$^2$, the CHOOZ experiment requires $\theta_{13} <
13^\circ$. We therefore take the view that a non-zero value of
$\theta_{13}$ can be at most of order $\epsilon$.
\footnote{Suppose a
zeroth-order texture gives $\theta_{13}^{(0)} \approx 1$. Since
perturbations may give $\theta_{E12} \approx 1$, it is conceivable
that on computing $V$ one finds a precise cancellation, so that
$\theta_{13} = 0$. In general such cases are fine tuned, and we do not
consider them further. For the textures which do give $\theta_{13}^{(0)} 
= 0$, it is important that the perturbations to $\lambda_E$ do not induce 
$\theta_{E12} \approx 1$, since this will in general lead to $\theta_{13}
\approx 1$.} 

\medskip

Many theories of flavor with hierarchical fermion masses yield small mixing 
angles --- as in the well-known example of the Cabibbo angle
$\theta_c \approx (m_d/m_s)^{1/2}$. Conversely, theories with large mixing angles
typically do not have mass hierarchies. To avoid this typical
situation, the textures for  $\lambda_{E,\nu}^{(0)}$ become highly
constrained. One must search either for a $\lambda_\nu^{(0)}$ which
gives a large $\theta_{\nu23}^{(0)}$,  while maintaining
the degeneracy $\Delta m_\odot^{2(0)} = 0$; or a
$\lambda_E^{(0)}$ which gives a large $\theta_{E23}^{(0)}$, while maintaining
the charged lepton mass hierarchy $m_{e,\mu}^{(0)} = 0$.\footnote{It 
has been noted~\cite{smallangle} that the atmospheric
data does not exclude a conventional interpretation, with small angles
following from mass hierarchies, because the relevant mass hierarchies
may not be large. With  $\theta_{E,\nu23}^{(0)}=0$, at order $\epsilon$
one typically finds $|\theta_{\nu23}| = (m_\odot/m_{{\rm atm}})^{1/2}$ and  
$|\theta_{E23}| = (m_\mu/m_\tau)^{1/2}$. For $\Delta m^2_\odot =
10^{-5}$ eV$^2$ and $\Delta m^2_{{\rm atm}} = 10^{-3}$ eV$^2$, and opposite signs 
for $\theta_{\nu23}$ and $\theta_{E23}$, one finds
$\sin^2 2 \theta_{23} = 0.82$, at the edge of the Super-Kamiokande 90\%
C.L. allowed region. It is very interesting that this case is not
excluded, but it is not favored, and we study the more revolutionary
case that at least one of $\theta_{\nu ,E23}^{(0)}$ is non-zero.
With a suitable see-saw structure $|\theta_{\nu23}| =
(m_\odot/m_{{\rm atm}})^{1/4}$ is also possible~\cite{fty}.}
There are few solutions to this puzzle, and
therefore few zeroth-order textures.

\paragraph{3}
In searching for textures for $\lambda_{\nu,E}^{(0)}$, we allow two
types of relations between otherwise independent, non-zero matrix
elements. 
\begin{itemize}
\item 
Any two entries of a matrix may be set equal. 
This can frequently result from a symmetry, but we do not
require that a symmetry origin can be found. 

\item
The determinant of each matrix, or any of its $2 \times 2$
sub-determinants, can be set to zero.
\end{itemize}
We do not allow other precise relations. The determinantal conditions
arise naturally when heavy states are integrated out, as in the
seesaw and Froggatt-Nielsen mechanisms. 
This may be due to zeroth order textures of the light-heavy
couplings for any (non singular) mass matrix of the heavy states, or to a
mass hierarchy of the heavy states themselves.
Suppose that a heavy charged
lepton couples to combinations $\Sigma_i a_i e_{Li}$ and  $\Sigma_i b_i e_{Ri}$
of left- and right-handed charged lepton flavor eigenstates. On
integrating out the heavy state, $\lambda_{Eij}^{(0)} \propto a_i
b_j$ and has only a single non-zero eigenvalue. Integrating out each
heavy state leads to a single eigenvalue in the light matrix. If there
is a hierarchy amongst the contribution of various heavy states, then
there will be a hierarchy of light eigenvalues. Only the dominant
contributions are included in $\lambda_{\nu,E}^{(0)}$, which may
therefore have zero determinants and sub-determinants.

\paragraph{4}
There are three zeroth-order neutrino mass matrix textures which satisfy~(\ref{eq:modot}),
(\ref{eq:theta23}) and (\ref{eq:theta13}) (with
$\theta \rightarrow \theta_\nu$), 
which we label $I$, $II$ and $III$: 
\begin{equation}
\lambda_\nu^{(0)I} = \pmatrix{0&B&A \cr B&0&0 \cr A&0&0} \hspace{0.5in}
\lambda_\nu^{(0)II} = \pmatrix{0&0&0 \cr 0&{B^2 \over A}&B \cr 0&B&A} 
\hspace{0.5in}
\lambda_\nu^{(0)III} = \pmatrix{A&0&0 \cr 0&0&A \cr 0&A&0}. 
\label{eq:3nu}
\end{equation}
Texture $I$ leads to a heavy pseudo-Dirac neutrino:
\begin{equation}
I: \; \; \mbox{Pseudo-Dirac} \hspace{0.5in} m_{\nu 1}^{(0)} =  m_{\nu 2}^{(0)}
 \; \gg \;  m_{\nu 3}^{(0)} = 0
\label{eq:I}
\end{equation}
while texture $II$ gives the zeroth-order hierarchical eigenvalue pattern:
\begin{equation}
II: \; \; \mbox{Hierarchical} \hspace{0.5in}
m_{\nu 3}^{(0)} \; \gg \;   m_{\nu 1}^{(0)} =  m_{\nu 2}^{(0)} = 0.
\label{eq:II}
\end{equation}
The third texture leads to complete degeneracy at zeroth-order:
\begin{equation}
III: \; \; \mbox{Degeneracy} \hspace{0.5in} m_{\nu 1}^{(0)} =  m_{\nu 2}^{(0)}
= m_{\nu 3}^{(0)} \neq 0.
\label{eq:III}
\end{equation}
Textures $I$ and $II$ can both be obtained by the seesaw mechanism. A simple model
for texture $II$ has a single heavy Majorana right-handed neutrino, $N$,
with interactions $l_{2,3}NH + MNN$, which could be guaranteed, for
example, by a $Z_2$ symmetry with $l_{2,3}, N$ odd and $l_1$ even.
A simple model for texture $I$ has two heavy right-handed neutrinos
which form the components of a Dirac state and have the interactions
$l_1 N_1 H + l_{2,3} N_2 H + M N_1 N_2$. These interactions could
result, for example, from a U(1) symmetry with $N_1, l_{2,3}$ having
charge +1, and $N_2, l_1$ having charge $-1$. In both cases, the missing 
right-handed neutrinos can be heavier, and/or have suitably suppressed 
couplings.

The search for charged lepton textures is very similar to that for neutrinos, 
except that the zeroth-order eigenvalues must be hierarchical, not 
pseudo-Dirac or degenerate.
There is only one zeroth-order charged lepton mass matrix texture which
satisfies~(\ref{eq:memu}), (\ref{eq:mtau}), (\ref{eq:theta23}) and
(\ref{eq:theta13}) (with $\theta \rightarrow \theta_E$) and
$\theta_{E12}^{(0)}=0$,
which we label $IV$
\begin{equation}
\lambda_E^{(0)IV} = \pmatrix{0&0&0 \cr 0&0&B \cr 0&0&A}
\label{eq:IV}
\end{equation}
which is the charged analogue of the hierarchical neutrino texture $II$;
we have rotated the right-handed charged leptons so that the
entries of the first two columns vanish.
Textures ($I$), ($II$) and ($IV$) have the special case $A=B$.

A special hierarchical
texture,
\begin{equation}
\lambda_E^{(0)V} = \pmatrix{A&A&A \cr A&A&A \cr A&A&A}
\label{eq:V}
\end{equation}
often considered in the literature~\cite{democratic},
gives $\theta_{13}^{(0)}=0$ only
when $\theta_{E12}^{(0)}$ takes the non-zero value corresponding to
the light eigenstates being $(1,1,-2)$ and $(1,-1,0)$, and is therefore an
exception to our usual requirement that $\theta_{E12}^{(0)} = 0$. In turn,
these two special directions will have to be picked out by the
perturbation which gives $m_\mu$. 
Note that texture $V$ is not
allowed  for the neutrino mass matrix, since it gives $\theta_{\nu13}^{(0)}
\simeq 35^\circ$, and clearly violates~(\ref{eq:theta13}).

\smallskip

Texture $II$ has been obtained by the seesaw mechanism by adding an extra 
singlet neutrino, or by adding $R$ parity violating interactions in 
supersymmetric theories~\cite{TextII}. The possibility of complete 
degeneracy via texture $III$ has also been noted~\cite{TextIII}. 
A search for theories giving the observed atmospheric and solar neutrino 
fluxes with three light neutrinos, but with more stringent assumptions than 
those adopted here, found both textures $I$ and $II$~\cite{bhssw}.
It is remarkable that all theories,
which explain both atmospheric and solar neutrino fluxes with 
three light neutrinos, must yield one of these zeroth-order textures.
Any exception must involve more complicated relations between matrix elements 
--- for example, with entries differing by a multiple or a sign~\cite{gg}.

\smallskip

Texture $III$ is unique: it is the only case where the zeroth-order
texture does not provide $\Delta m_{{\rm atm}}^2$ (for neutrinos) or
$m_\tau$ (for charged leptons).
Apparently this texture does not guarantee a large value for $\theta_{23}$:
at zeroth-order the three neutrinos are all degenerate, 
so that degenerate perturbation theory may lead to a large value for
$\theta_{23}^{(1)}$, perhaps cancelling $\theta_{23}^{(0)}$.
This is not correct --- small perturbations to a Dirac mass
term yield a pseudo-Dirac neutrino with a mixing
angle close to $45^\circ$ --- texture ($III$) yields the
result $\theta_{23}^{(0)} = 45^\circ$, and this receives only small
corrections from higher order. This is to be compared with the Super-Kamiokande
data: $\theta_{23} = 45^\circ \pm 13^\circ$ at 90\% C.L.

\paragraph{5}
In listing textures $I$--$V$, we have not discussed the form of the
lepton mass matrix which is not explicitly displayed. A neutrino
texture from~(\ref{eq:3nu}) could be paired with a diagonal charged
lepton mass matrix, or with texture $IV$. 

It is straightforward to find symmetries which yield textures $I$ and
$II$; an example of $L_e-L_\mu-L_\tau$ was given in~\cite{bhssw}. Any
symmetry yielding these textures does not distinguish $l_\mu$ from $l_\tau$, 
and hence requires both of
these textures to be paired with the charged lepton mass matrix of
texture $IV$. We find two zeroth-order schemes for lepton masses:
\begin{eqnsystem}{sys:s}
\mbox{Scheme~A}&& \lambda_\nu^{(0)I} \oplus \lambda_E^{(0)IV}
\label{schemeA}\\
\mbox{Scheme~B}&& \lambda_\nu^{(0)II} \oplus \lambda_E^{(0)IV}.
\label{schemeB}
\end{eqnsystem}
For these schemes the special case $A=B$, in both charged and neutral 
matrices, gives $\theta_{23}^{(0)}=0$, and is not allowed. Even if $A=B$ in 
just one of the matrices, neither scheme can claim to predict $\theta_{23}$ 
near $45^\circ$, as suggested by the Super-Kamiokande data.

We do not know how textures $III$  can be obtained from symmetry arguments. 
The democratic texture $V$ follows from the symmetry group $S_{3L}
\times S_{3R}$~\cite{democratic}. With $\ell_L$ $(\ell_R)$ transforming as
$1_L \oplus 2_L$ $(1_R \oplus 2_R)$ of $S_{3L}(S_{3R})$,
$\lambda_E^{(0)} = A (1_L 1_R)$, which is the democratic form of~(\ref{eq:V}).
However, a scheme for both charged and neutral lepton masses does not
result at zeroth order. The neutrino mass matrix involves two
invariants: $\lambda_\nu^{(0)} = B (1_L 1_L) + C (2_L 2_L)$, and is
diagonalized by the rotation matrix $V_\nu^{(0)} = V_e^{(0)}$, giving
$V^{(0)} = I$. Even if $B$ and $C$ are fine-tuned, or a suitably extended
symmetry is spontaneously broken in a special direction~\cite{caz}, so that
$\lambda_\nu^{(0)} \propto I$, the lepton mixing angles are then all
determined by the higher order perturbations.

The only zeroth-order schemes for lepton masses which we can 
justify from symmetries are $A$ and $B$. In scheme $A$ the solar neutrino 
oscillations are large angle , while in $B$ the angle may be large or small,
and is entirely determined by the perturbations. Neutrino hot dark matter 
and a $\beta \beta_{0\nu}$ signal are possible only in texture $III$, and
hence do not occur in either scheme.

\smallskip

Models for the zeroth-order schemes $A$ and $B$, in which the neutrino
masses arise from the seesaw mechanism, are easy to construct, and can
also be extended to SU(5) unification.
Scheme $B$ occurs in theories with a single heavy Majorana
right-handed neutrino, $N$, and any flavor symmetry with $N, l_{2,3},
e_3, H_{u,d}$ transforming trivially, but $l_1, e_{1,2}$ transforming
non-trivially such that $l_1 e_{1,2} H_d$ is forbidden. An example is
provided by a U(1) symmetry with $l_1, e_{1,2}$ positively charged.
This can be extended to a theory with three right-handed neutrinos by
introducing $N_1$ and $N_2$, with equal and opposite charges, such
that the only allowed renormalizable operator is the mass term 
$M' N_1 N_2$.

Scheme $A$ occurs in a theory which has two heavy right-handed
neutrinos,
which form the components of a Dirac state, and the interactions
$l_1 N_1 H_u + l_{2,3} N_2 H_u + M N_1 N_2 + l_{2,3} e_3 H_d$.
For example, this follows from a U(1) flavor symmetry with
$N_1, l_{2,3}$ having charge +1, $N_2, l_1, e_{1,2}, H_d$ having charge $-1$,
$H_u$ and $e_3$ being neutral.
This can be extended to a theory with three right-handed neutrinos by
introducing the neutral state $N_3$.

In our example $U(1)$ theories, for both schemes $A$ and $B$, in the trivial 
extensions to $SU(5)$ where each field is replaced by its parent $SU(5)$ 
multiplet ($e_3 \rightarrow T_3$ etc), the only additional invariant 
interaction is $T_3 T_3 H_u$, yielding the top quark Yukawa coupling at 
zeroth order.

\paragraph{6}
We have studied textures for lepton mass matrices with three light
neutrinos, which at zeroth-order give a large mixing angle
for the oscillation $\nu_\mu \rightarrow \nu_\tau$, and a vanishing 
mixing angle for $\nu_\mu \rightarrow \nu_e$. We allow textures with
zero determinants and sub-determinants, and we allow independent
entries to be set equal.
In the case of the neutrino mass matrix, requiring two degenerate
neutrinos to allow $\Delta m_\odot^2 \ll \Delta_{{\rm atm}}^2$, there are 
just three such zeroth-order textures; while in the charged case,
requiring a zeroth-order mass for $m_\tau$ but not for $m_\mu$ or
$m_e$, there are only two possibilities. For a flavor symmetry to
yield both charged and neutral textures with these zeroth-order
properties, only two schemes are possible, both of which can be
obtained in simple seesaw models.

Many other theories of neutrino masses exist which can account for
both atmospheric and solar neutrino fluxes.  They involve more than
three light neutrinos, textures with entries related in some special
way, or theories where $m_\tau, \theta_{23}$ and/or $\theta_{13}$ are
determined only at higher order. However, the zeroth-order textures
and schemes which we have found appear particularly simple.

\paragraph{Acknowledgements}
We thank Graham Ross and Andrea Romanino for useful discussions.
This work was supported in part by the
U.S. Department of Energy under Contracts DE-AC03-76SF00098, in part
by the National Science Foundation under grant PHY-95-14797.

\end{document}